\documentclass{sig-alternate-nocopyright}
\usepackage{epsfig}
\usepackage{enumerate}
\usepackage{graphicx}

\usepackage{pgfplots}
\pgfplotsset{compat=1.3}
\pgfplotscreateplotcyclelist{experimentbar}{%
black,fill=yellow,mark=none\\%
black,fill=blue!30!white,every mark/.append style={fill=cyan!80!black},mark=none\\%
black,fill=red!60!white,every mark/.append style={fill=orange!80!black},mark=none\\%
black,fill=teal!60!white,every mark/.append style={fill=lime},mark=none\\%
black,fill=orange!60!white,every mark/.append style={solid,fill=gray},mark=none\\%
black,fill=gray!50!white,every mark/.append style={solid,fill=yellow!80!black},mark=none\\%
black,fill=blue!60!white,every mark/.append style={fill=teal!80!black},mark=none\\%
black,fill=brown,every mark/.append style={solid,fill=red!80!black},mark=none\\%
}

\newcommand{\stitle}[1]{\vspace{0.8ex} \noindent{\bf #1}}
\newcommand{\eat}[1]{}

\author{\selectfont\rmfamily\itshape
Technical Report\\
\selectfont\ttfamily\itshape
August 2, 2013}

\title{Hyper-Graph Based Database Partitioning for Transactional Workloads}

\begin{document}
\maketitle

\pagestyle{plain}
\setcounter{page}{1}
\pagenumbering{arabic}

\begin{abstract}
A common approach to scaling transactional databases in
practice is horizontal partitioning, which increases system scalability,
high availability and self-manageability. Usually it is very
challenging to choose or design an optimal partitioning scheme for
a given workload and database. In this technical report, we propose
a fine-grained hyper-graph based database partitioning system for
transactional workloads. The partitioning system takes a database,
a workload, a node cluster and partitioning constraints as input and
outputs a lookup-table encoding the final database partitioning decision.
The database partitioning problem is modeled as a
multi-constraints hyper-graph partitioning problem. By deriving a
min-cut of the hyper-graph, our system can minimize the total number
of distributed transactions in the workload, balance the sizes
and workload accesses of the partitions and satisfy all the partition
constraints imposed. Our system is highly interactive as it allows
users to impose partition constraints, watch visualized partitioning
effects, and provide feedback based on human expertise and indirect
domain knowledge for generating better partitioning schemes.
\end{abstract}

\section{INTRODUCTION}


The difficulty of scaling front-end applications is well known for DBMSs executing highly concurrent workloads.
One approach to this problem employed by many Web-based companies is to partition the data and workload across
a large number of commodity, shared-nothing servers using a cost-effective, parallel DBMS, e.g. Greenplum Database.
The scalability of online transaction processing (OLTP) applications on these DBMSs depends on the existence of
an optimal database design, which defines how an application's data and workload is partitioned across nodes in
the cluster, and how queries and transactions are routed to nodes. This in turn determines the number of transactions
that access data stored on each node and how skewed the load is across the cluster. Optimizing these two factors is
critical to scaling complex systems: a growing fraction of distributed transactions and load skew can degrade
performance by over a factor 10x. Hence, without a proper design, a DBMS will perform no better
than a single-node system due to the overhead caused by blocking, inter-node communication,
and load balancing issues.

\eat{
Transactional workloads, featured with numerous short-lived and highly concurrent transactions,
a small set of pre-defined transaction types and relatively few tuples touches by each transaction,
are very different from analytical workloads.
In practice, the most common technique used to scale transactional databases is horizontal partitioning.
By distributing the partitions across a number of physical nodes and routing the queries to the corresponding nodes,
both throughput and scalability are increased.
}

Usually, it is very challenging to choose or design an optimal partitioning scheme for a given workload and database.
Executing small distributed transactions will incur heavy overhead~\cite{schism} and thus should be avoided whenever possible.
However, especially when dealing many-to-many relationships or very complex database schemas,
it is not an easy task to put all the tuples that are accessed together onto the same node so as to
reduce the overhead of distributed transactions.
In the meantime, data skew or workload skew degrades the performance of the overloaded nodes
and thereby lowers the overall system throughput.
Therefore, it is also very critical to achieve both data and workload balancing.
Moreover, for a specific partitioning strategy to be feasible,
it must not violate the constraints on the cluster configuration,
such as node storage capacity, node processing ability, and network bandwidth between nodes.

Partitioning in databases has been widely studied, for both single system servers and
shared-nothing systems. However, most of the existing techniques for automatic database
partitioning are tailored for large-scale analytical applications (i.e. data warehouses).
These approaches typically produce possible partitions using round-robin (send each
successive tuple to a different partition), range (divide up tuples according to a set of
predicates), or hash-partitioning (assign tuples to partitions by hashing them)~\cite{paradb}, which
are then evaluated using heuristics and cost models. Unfortunately, none of these approaches
are ideal for transactional workloads, which are very different from analytical
workloads and are featured with numerous short-lived and highly concurrent transactions,
a small set of pre-defined transaction types and relatively few tuples touches by each transaction.
For transactional workloads, if more than one tuple is accessed, then
round-robin and hash partitioning typically require accessing to multiple sites and thus
incur distributed transactions, which as we explained have significant overhead. Range
partitioning may be able to do a better job, but this requires carefully selecting ranges
which may be difficult to do by hand. The partitioning problem gets even harder when
transactions touch multiple tables, which need to be divided along transaction boundaries.
For example, it is difficult to partition the data for social networking web sites,
where schemas are often characterized by many n-to-n relationships.

In this report, we introduce a fine-grained hyper-graph
based database partitioning system for transactional workloads. The
input to our system includes a database, a workload, a node cluster
and partitioning constraints imposed by users. We model the
database partitioning problem as a multi-constraints hyper-graph
partitioning problem. Our system first analyzes the database
and workload and constructs a weighted hyper-graph. It then runs
an iterative hyper-graph partitioning phase to get a feasible and
near-optimal partitioning scheme. After each iteration of partitioning,
our system will evaluate the partitioning feasibility and performance,
receive user feedbacks and then decide whether it should
do hyper-graph refinement and re-partitioning. The final output is
a lookup table which indicates how the database should be partitioned
and distributed over the cluster so that the total distributed
transactions in the workload will be minimized, the sizes and workload
accesses of the partitions will be balanced and all the imposed
constraints will be met.

Our database partitioning system can easily handle
many-to-many table relationships and complex database schemas. It is
also efficient as the size of the derived hyper-graph is independent
of the database size. It provides great opportunities for the users
to participate in the loop of decision making and import their human
expertise and indirect domain knowledge for better partitioning
performance.

The rest of the report is organized as the follows:
Section~\ref{sec:model} describes the hyper-graph based database partitioning model.
Section~\ref{sec:architecture} presents the partitioner system architecture, as well as implementation details.
Section~\ref{sec:experiment} introduces the experiments evaluation.
Section~\ref{sec:relatedwork} is the related works.
We conclude in Section~\ref{sec:conclusion}.

\section{Hyper-Graph Based Database Partitioning}
\label{sec:model}

Here we focus on horizontal partitioning of database tables.
The effect of a partitioning scheme for a transactional workload is normally measured by the number of distributed transactions~\cite{schism}.
So the problem can be turned into finding a partitioning scheme that minimizes the number of distributed transactions.
Data skew and workload skew will decrease the system throughput and thus are expected to be under certain threshold.
There are also constraints imposed for the partitioning in practice,
such as node storage capacity, node processing ability and network bandwidth between physical nodes.
For a partitioning strategy to be feasible, it must meet all these constraints.
We thereby formalize the database partitioning problem as follows:
\vskip 5pt
{\em Given a database D, a workload W, the number of physical nodes k, and the constraints C,
find the optimal partitioning solution to partition D over k physical nodes
so that the cost of executing W is minimized,
while all the constraints C are satisfied and the imbalance degree of the data sizes
and workload accesses across k nodes are under some balance threshold T.}
\vskip 5pt

\stitle{Tuple Group}. Before modeling the above database partitioning problem as a multi-constraints hyper-graph partitioning problem,
we first give the definition of {\em tuple group}.

A tuple group is a collection of tuples within a relation, which will always be accessed together throughout the execution of $W$.
Each tuple group is essentially represented by a {\it min-term predicate}~\cite{pdds}.
Given a relation $R$, where $A$ is an attribute of R,
then a simple predicate $p$ defined on R has the form
$$p : A\ \theta\ const$$
where $const$ is a constant value and $\theta \in \{=,<,\neq,>,\leq,\geq\}$.

A min-term predicate is the conjunction of simple predicates.
Given the set of simple predicates $\{p_{1},p_{2},...,p_{n}\}$ on relation $R$ that are derived from $W$,
a min-term predicate $M$ is defined as
$$M = p_1^* \wedge p_2^* ... \wedge p_n^*$$
where $p_{i}^*=p_{i}$ or $p_{i}^*=\neg p_{i}$ ($1 \leq i \leq n$), which means that each simple predicate can occur in a min-term
predicate either in its natural form or its negated form.

The min-term predicate has the property that all the tuples belonging to this predicate will be accessed together.
A min-term has two attributes: min-term size and access count.
The min-term size is the number of tuples it represents in the actual table.
The access count is the times that transactions within the workload accessing (some of) the tuples covered by this min-term predicate.
These two attributes of a tuple group $M$ are denoted by $size(M)$ and $access(M)$ respectively.

\stitle{Hyper-Graph Partitioning Problem Modeling}.
It is obvious that a good partitioning scheme should put all the tuples of a tuple
group into the same node in order to reduce the number of distributed transactions.
So our basic idea to do the partitioning is: we first analyze and split $D$ into disjoint tuple groups,
then try to place these tuple groups into $k$ nodes.

A hyper-graph extends the normal graph definition so that an edge can connect any number of vertices.
A hyper-graph $HG(V,E)$ is constructed as follows:
each vertex $v_i$ represents a tuple group $M_i$;
each hyper-edge $e_i=(v_1,v_2,...,v_n)$ represents a transaction $X_i$ in $W$
accessing all the tuple groups connected by this hyper-edge.
A vertex $v_i$ has two kinds of weights $size(M_i)$ and $access(M_i)$.
The weight $count(e_i)$ of a hyper-edge $e_i$ is the number of transactions that access the same vertices (i.e. tuple groups).

Given a hyper-graph $HG(V,E)$, k-way partitioning of $HG$ assigns vertices $V$ of $HG$ to $k$ disjoint nonempty partitions.
The k-way partitioning problem seeks to minimize the net cut, which means the number of hyper-edges that
span more than one partition on the graph partitioning, or, more generally, the sum of weights of such hyper-edges.
There are also constraints imposed on the graph partitioning, which correspond to the partition constraints $C$ and the balance threshold $T$
in the above database partitioning problem.

Each cut-edge incurs at least one distributed transaction
since the data that the transaction need to access will be placed into at least two nodes.
So the sum of weights of the cut-edges is equal to the total number of resulting distributed transactions.

As such, we turn the database partitioning problem into a multi-constraints hyper-graph partitioning problem
which aims to get the minimum k-way net-cut while keeping graph partitions balanced and meeting various constraints.

\eat{
\subsection{Skew Factor}
We define a skew factor $SF$ to quantitatively
measure the extent of data and workload skews. Assume a cluster with n nodes.
Let $s_i$ and $t_i$ be the size of assigned database partition and the number of accessing
transactions respectively, of the $i$th node. Then $SF$ is calculated as follows:
$$SF=\frac{\sum\limits^n_{i=1}(\alpha \times(s_i-\frac{1}{n}\times \sum\limits^n_{i=1}s_i)^2+\beta \times(t_i-\frac{1}{n}\times\sum\limits^n_{i=1}t_i)^2)}{n}$$
where $\alpha$ and $\beta$ are configurable non-negative parameters which may be used to reflect the different
performance impacts of data skew and workload skew. $\alpha + \beta = 1$. Generally, a smaller value of
$SF$ means a better partitioning result.

In addition, we define the data skew factor $DSF$ and workload skew factor $WSF$ separately.
For users who only care for the data skew, we set $\beta=0$, then $$DSF=\frac{\sum\limits^n_{i=1}(s_i-\frac{1}{n}\times \sum\limits^n_{i=1}s_i)^2}{n}.$$
In the similar, if we set $\alpha=0$, then the workload skew factor $$DSF=\frac{\sum\limits^n_{i=1}(t_i-\frac{1}{n}\times \sum\limits^n_{i=1}t_i)^2}{n}.$$
}

\section{SYSTEM Description}
\label{sec:architecture}
We first introduce the overview system architecture, and then present
the implementation details.

\begin{figure}[htb!]
\centering
\includegraphics[height=2.2in]{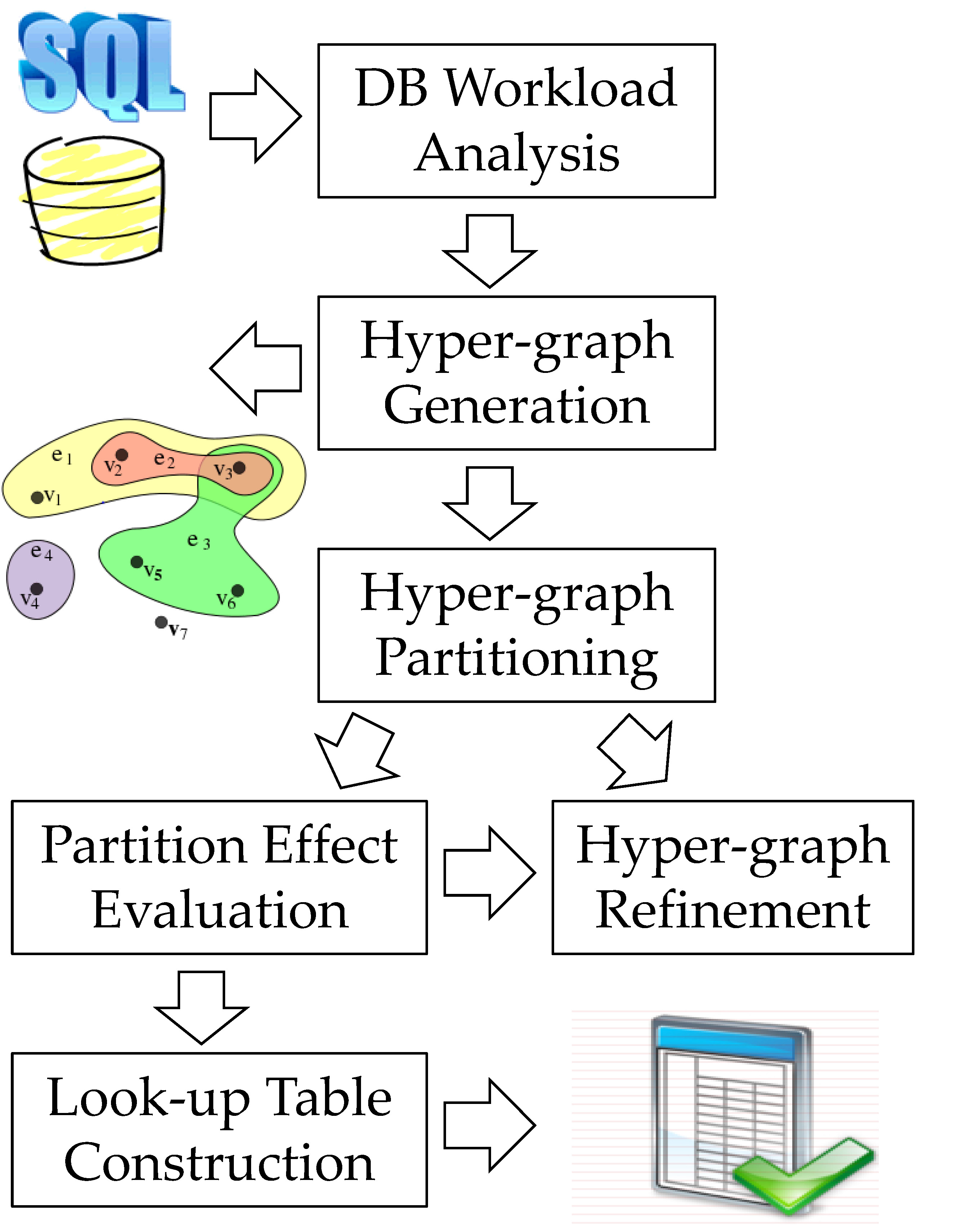}
\caption{System Architecture}
\label{fig:framework}
\vskip -20pt
\end{figure}

\subsection{System Overview}

Figure~\ref{fig:framework} illustrates the overview of our solution, which consists of the following six steps:

{\bf S1: DB and workload analysis.} Each table in the database is divided into one or multiple
{\it tuple groups}, according to the information extracted from the workload. The tuples within each group are always
accessed together throughout the whole workload. The sizes of tuple groups are derived from the database meta-data
and statistics stored in the system catalog. Besides, the information about which tuple groups are involved in each
transaction of the workload is also recorded.

{\bf S2: Hyper-graph generation.} The database partitioning problem is modeled as a hyper-graph partitioning problem.
The hyper-graph has the following characteristics: 1.~each tuple group obtained from the previous step corresponds to
a distinct graph vertex with two weights: the tuple group size and the number of transactions accessing this tuple group;
2.~each transaction is mapped to a hyper-edge that connects all the tuple groups it accesses. It is possible for
different transactions to be mapped to the same hyper-edge. Each hyper-edge is associated with a weight counting the
number of transactions mapped to it.

{\bf S3: Hyper-graph partitioning.} A graph partitioning algorithm is used to produce a balanced min-cut partitioning
of the hyper-graph into $k$ partitions. Each vertex (i.e. tuple group) is assigned to one partition, and each partition
is assigned to one cluster node. The min-cut of the hyper-graph means a minimized number of distributed transactions
resulting from the corresponding database partitioning strategy. The partitioning algorithm also tries to keep the extent of
incurred data skew and workload skew under certain thresholds.

{\bf S4: Partitioning effect evaluation.} The graph partitioning result from S3 is evaluated according to certain
criteria. If the result meets the criteria, the next step is {S6}; otherwise it is {S5}. The criteria are
two-fold. First, the resulting database partitioning must be feasible, which means that it should not violate
the physical constraints of the cluster. For example, the total volume of data assigned to a cluster node must not exceed
its storage capacity. Second, the partitioning performance, i.e. the number of distributed transactions and the
extent of data skew and workload skew, should achieve the expectations that are optionally imposed by the user.
During this phase, user can watch the visualized partitioning effects, and optionally provide feedback based on
his expertise and domain knowledge to affect the decision on whether the system should proceed to do graph refinement and re-partitioning.

{\bf S5: Hyper-graph refinement.} The existing hyper-graph is refined towards generating a better
partitioning that meets the criteria defined in S4. The basic idea of refinement is to choose some
tuple groups in the hyper-graph and break them into smaller ones as new vertices. The hyper-edges are adjusted accordingly.
The newly derived hyper-graph is then fed into S3 for partitioning.
Intuitively, the new hyper-graph represents an expanded solution space that subsumes the space represented
by the old hyper-graph.
Since the new hyper-graph is usually similar to the old one, in addition to running the complete partitioning algorithm,
the partitioning of the former could be done by incrementally revising the partitioning result of the latter.

{\bf S6: Look-up table construction.} The finally decided database partitioning strategy is encoded into a look-up table, which
records the tuple-to-node mappings via a compact data structure representation. This look-up table is used when both loading the database
into the cluster and routing transactions to involved data nodes during workload execution.

\vspace{2mm}
In the following sections, we elaborate on the technical details of our database partitioning
solution roughly depicted above.

\subsection{DB and Workload Analysis}

The steps for obtaining the tuple groups, i.e. min-term predicates, for each relation $R$ are illustrated bellow:
First, extract all the simple predicates related to relation $R$ in the workload.
Second, construct the min-term predicates list by enumerating the conjunctions of all the simple predicates of either normal or negated form.
Third, eliminate those min-term predicates containing contradicting simple predicates,
and simplify the min-term predicates by removing the simple predicates that are implied by other simple predicates within the same min-term predicate.
In order to control the number of min-term predicates generated, we could only select the top-$k$ mostly accessed attributes of each relation for min-term predicate construction.
$k$ is configurable by the user and currently has a default value of 2.

We obtain the database meta-data and statistics information (e.g. histograms) from the underlying database system catalog, and then estimate $size(M)$ of a min-term predicate with methods similar to those utilized by a conventional relational database optimizer.
To obtain the access count $access(M)$ of a min-term predicate,
we examine each transaction in the workload and determine whether it accesses the tuple group $M$.
A transaction $X$ will access the tuple group $M$ iff for each attribute $A$ of $R$, the set of
simple predicates on $A$ that are involved by $X$ don't contradict with $M$.
Then $access(M)$ is equal to the total number of transactions accessing the tuple group $M$.

The outputs of the DB and workload analysis include the min-term predicates for all the database relations w.r.t the workload,
and a transaction access list which tells which min-term predicates a transaction will access.

\subsection{Hyper-Graph (Re-)Partitioning}

Our partitioning system employs an existing partitioning algorithm hMETIS~\cite{hmetis} to do the
hyper-graph partitioning. hMETIS is the hyper-graph version of
hMETIS, a multilevel graph partitioning algorithm.
hMETIS will tell which vertex belongs to which node, and also the sum of weights
of the net cut, which represents the number of distributed transactions
that would be incurred by this partitioning solution.

hMETIS also supports
incrementally revising an already partitioned hyper-graph according to new constraints.
This feature of hMETIS enables the lighter-weight hyper-graph repartitioning after the
hyper-graph refinement.

\subsection{Partitioning Effect Evaluation}

\subsubsection{Partitioning Effect Criteria}

For a specific partitioning solution to be feasible, it must not violate the physical restrictions of the underlying node cluster.
Three types of physical restrictions are considered. First, the storage capacity of each node is limited. Second, the data processing
ability of each node, which depends on the CPU and I/O speeds, is also limited. Third, the bandwidths of the network connecting the nodes
are limited.

Intuitively, when a node is assigned more data and accessed by more transactions, the speed at which this node handle transaction processing will be
slower, thus this node is more likely to become a performance bottleneck of the whole system.
Therefore, the extent of data skew and workload skew of the system resulting from a specific partitioning solution should be within certain threshold
which represents the performance expectation of the user.
We define a skew factor $SF$ to quantitatively measure the extent of data and workload skews. Assume a cluster with $n$ nodes.
Let $s_i$ and $t_i$ be the size of assigned database partition and the number of accessing transactions respectively, of
the $i$th node. Then $SF$ is calculated as follows:
$$SF = \frac{\sum\limits_{i=1}^{n} (\alpha \times (s_i - \frac{1}{n} \times \sum\limits_{i=1}^{n} s_i)^2 +
\beta \times (t_i - \frac{1}{n} \times \sum\limits_{i=1}^{n} t_i)^2)}{n}$$
\noindent where $\alpha$ and $\beta$ are configurable non-negative parameters ($\alpha + \beta = 1$) which may be used to reflect the different performance
impacts of data skew and workload skew.
Generally, a smaller value of $SF$ means a better partitioning result.

Finally, the user also inputs his expected number of partitioning iterations (i.e. the cycle of S3 $\rightarrow$ S4 $\rightarrow$ S5 $\rightarrow$ S3 in Figure~\ref{fig:framework}),
which represents the time budget that the user allows the system to consume before he gives up finding a feasible or better partitioning result.

\subsubsection{Evaluation Report and User Interaction}

The evaluation generates predictions on multiple performance metrics: data distribution, workload distribution,
the number of distributed transaction, as well as the system throughput and response latency, which
are obtained by the simulated execution of the workload with our previous tool PEACOD~\cite{peacod},
a partitioning scheme evaluation and comparison system.

Our system is always under one of two
execution modes: the {\it fully automatic mode} and the {\it interactive mode}. The automatic mode will totally
rely on the intelligence of the system on evaluating the partitioning performance in order to decide whether continue
or halt the iterative graph partitioning procedure. In contrast, the interactive mode allows the
user to provide feedback to affect the partitioning strategy at runtime, jointly with the system intelligence.

Under the interactive mode, after each graph partitioning iteration, the system will produce the visualized partitioning results.
Besides, the comparison on the partitioning results between this and the last iteration will also be visualized.
After that, the system will pause execution and wait for the instructions or feedback from the user.
The interactions by the user can be of various types. First, the user may terminate the system execution earlier, with either an already satisfactory partitioning
result or a hopelessly bad result. Second, the user may
provide suggestions on how the current hyper-graph should be refined.

\eat{
The performance report contains the following metrics: data distribution, workload distribution,
number of distributed transaction, and maybe throughput and response time.
The key metric is the number of distributed transaction.
Partition Evaluator chooses the best one mainly according to this metric.
Data distribution and workload distribution are the balance factors of data size and access frequency, which is the reference to user.
Throughput and response time are also obtained by simulated executing the given workload by our previous tool PEACOD~\cite{peacod},
which is a partitioning scheme evaluation and comparison tool.

The performance report will exhibit and interact with user.
User can determine whether to do partitioning refinement by two ways.
The one is to provide the parameter $\#iter$ to let our partitioning system automatically do.
The other is that user can examine the performance result and determine whether he is satisfied.
If not, he can let the partitioning system to do refinement until he satisfies the result.
}

\subsection{Hyper-Graph Refinement}
If the partition result is neither feasible nor good enough, we
invoke the partitioning refinement to get a feasible and better one.
The basic idea is to split some tuple groups (i.e. hyper-graph vertices) and then redo partitioning
for the according revised hyper-graph. Tuple group splitting is three-phase.

First, we rank the vertices with a ranking function.
Vertices with higher ranks are more likely to be split.
Currently, we use the vertex size as the ranking function.
Alternative rank functions, e.g. the ratio of size and access frequency, may also be utilized.

Second, we select the top-$k$ vertices to split. $k$ is configurable by the user and currently has default value of 20.

Last, we split each selected vertex $V$ into two new vertices $V_1$ and $V_2$.
We pick up the simple predicate $p$ with the lowest selectivity in the min-term predicate $M$ of $V$ and then
break $p$ into two simple sub-predicates, $p_1$ and $p_2$, with the same selectivity. $V_1$ and $V_2$ correspond to
the new min-term predicates constructed by replacing $p$ in $M$ with $p_1$ and $p_2$ respectively.
A hyper-edges accesses $V_1$ and $V_2$ iff it accesses $V$. As a result,
$size(V_1) = size(V_2) = size(V)/2$ and $access(V_1) = access(V_2) = access(V)$.

Obviously, hyper-graph refinement through splitting vertices can't further reduce the number of distributed transactions.
However, the refined hyper-graph does contain finer-grained vertices, which may enable feasible partitioning solutions as well
as mitigate the issues of data and workload skews.

\section{EXPERIMENTS EVALUATION}
\label{sec:experiment}
In this section, we report the experimental results.

\subsection{Environment Setup}
We have implemented a tool called PEACOD~\cite{peacod} to automatically and extendibly evaluate and compare
various database partitioning schemes. PEACOD is a Java application runs on Linux system.
The tool embeds several well-known OLTP benchmarks such as TPC-C, EPINIONS, TATP.
We shall make PostgreSQL~\cite{postgresql} as the target database server.

The experiments used PostgreSQL 9.1.2 as the DBMS with buffer pool size set to 1GB,
hosted on a machine with two 2.4GHz cores and 4GB of physical RAM.

\subsection{Partitioning Schemes}
We have implemented and embedded seven partitioning schemes to be compared,
including our hyper-graph based partitioning scheme({\bf HGP}).
The other six schemes are:

%

\vspace{2mm}
\stitle{CountMaxRoundRobin(CMRR).} In this scheme, those most frequently accessed attributes are selected
as the partitioning keys. The tables are partitioned in the round-robin manner based on the partitioning
key values.

\stitle{SchemaHashing(SH).} This scheme selects partitioning keys based on the primary-foreign key relationship
topology in the database schema~\cite{elastras}. The primary key of root table becomes the main
deriving partitioning key. Then the tables are hash-partitioned.

\stitle{PKHashing(PKH).} This scheme selects primary keys of tables as the partitioning keys and hash-partitions tables.

\stitle{PKRange(PKR).} This scheme selects primary keys of tables as the partitioning keys and range-partitions tables.

\stitle{PKRoundRobin(PKRR).} This scheme selects primary keys of tables as the partitioning keys and partitions tables
in the round-robin manner based on the partitioning key values.

\stitle{AllReplicate(AllR).} This scheme replicates each tables to all data nodes.

\subsection{Benchmarks}
In the experiment, we used the following three transactional benchmarks:

\vspace{2mm}
\stitle{TPC-C.} This benchmark is the current industry standard to evaluate the performance of
OLTP systems~\cite{tpcc}. It consists of nine tables and five transactions that simulate a warehouse-centric
order processing application. All the transactions are associated with a parameter warehouse id,
which is the foreign key ancestor for all tables except ITEM table. In the experiments, we generate
a 2-warehouse dataset and a 10-warehouse dataset.

\stitle{EPINIONS.} The Epinions.com experiment aims to challenge our system with
a scenario that is difficult to partition. It verifies it effectiveness in
discovering intrinsic correlations between data items that are not
visible at the schema or query level. It consists of four tables: users, items, reviews and trust.
The reviews table represents an n-to-n relationship between
users and items (capturing user reviews and ratings of items). The
trust table represents a n-to-n relationship between pairs of
users indicating a unidirectional trust value.
The workload is obtained from the open-source OLTP benchmarks oltpbenchmark~\cite{oltpbench}.

\stitle{TATP.} This benchmark is an OLTP testing application that simulates a typical caller location
system used by telecommunication provider~\cite{tatp}. It consists of four tables, three of which are
foreign key descendants of the root SUBSCRIBER table. Most of the transactions in TATP are associated
with SUBSCRIBER id, allowing them to be routed directly to the correct node.

\subsection{Number of Distributed Transactions}
In this experiments, we evaluate the number of distributed transactions that each scheme will produce.
We regard the key metric of a partitioning scheme is the number of distributed transactions.
We did several experiments to get the number for the 7 partitioning schemes and 3 benchmarks we mentioned above.

\subsubsection{TPC-C}

\begin{figure} \centering
\vskip -5pt
\begin{tikzpicture}[scale=0.8]
    \begin{axis}[ 
        ylabel=\# distributed transactions,
        y label style={at={(-0.08,0.50)},anchor=near ticklabel},
        height=6cm,
        width=9cm,
        grid=major,
        ybar=0pt,
        bar width=6pt,
        enlarge x limits=0.15,
        symbolic x coords={2w-2,10w-2,10w-5,10w-10},
        xtick=data,
        extra y ticks={250,750},
        legend style={at={(1.01,0.50)},anchor=west,legend columns=1,draw=none,font=\small},
        legend cell align=left,
        cycle list name=experimentbar,
    ]
    \addplot
    coordinates {(2w-2,75) (10w-2,32) (10w-5,59) (10w-10,82)};
    \addplot
    coordinates {(2w-2,65) (10w-2,135) (10w-5,159) (10w-10,168)};
    \addplot
    coordinates {(2w-2,246) (10w-2,234) (10w-5,377) (10w-10,418)};
    \addplot
    coordinates {(2w-2,142) (10w-2,230) (10w-5,377) (10w-10,417)};
    \addplot
    coordinates {(2w-2,260) (10w-2,241) (10w-5,368) (10w-10,410)};
    \addplot
    coordinates {(2w-2,570) (10w-2,547) (10w-5,821) (10w-10,914)};
    \addplot
    coordinates {(2w-2,930) (10w-2,927) (10w-5,927) (10w-10,927)};
    \legend{HGP,SH,PKH,PKR,PKRR,CMRR,AllR}
    \end{axis}
\end{tikzpicture}
\vskip -10pt
\caption{\# Distributed Transactions of TPC-C}
\label{fig:tpcc}
\vskip -10pt
\end{figure}
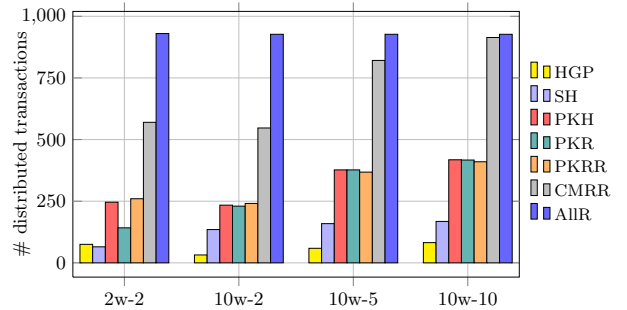

We first conducted the experiments using the TPC-C benchmark.
All the experiments used 1000 transactions workload.
There are three scenarios we tested:
partitioning a 2-warehouse-dataset into 2 nodes,
partitioning a 2-warehouse-dataset into 8 nodes,
and partitioning a 10-warehouse-dataset into 10 nodes.
We tested all the 7 partitioning schemes.
The result is listed in Table \ref{table:numdist} and Figure \ref{fig:tpcc}.

\begin{table}[ht]
\centering
\resizebox{\columnwidth}{!}{%
\begin{tabular}{c c c c c c c c}
\hline\hline
\  & HGP & SH & PKH & PKR & PKRR & CMRR & AllR  \\
\hline
2w->2 & 75 & 65 & 246 & 142 & 260 & 570 & 930 \\
10w->2 & 32 & 135 & 234 & 230 & 241 & 547 & 927 \\
10w->5 & 59 & 159 & 377 & 377 & 368 & 821 & 927 \\
10w->10 & 82 & 168 & 418 & 417 & 410 & 914 & 927 \\
\hline
\end{tabular}
}
\caption{\# Distributed Transactions for TPC-C}
\label{table:numdist}
\end{table}

From the result, we can observe that HGP and SH is significantly better than other partitioning schemes.
AllR is worst since it needs lots of update operation spanned over all physical nodes.
CMRR is also very bad since it chooses bad partitioning key.
The result of the three primary key based partitioning schemes are not very bad since they choose the best partitioning keys.
The most suitable partitioning keys of TPC-C are the primary keys for each table.
But it is not right for all OLTP benchmarks. So we can find that these
three partitioning schemes in other benchmark may perform very bad in the following experiments.

SH chooses the optimal partitioning keys according to the PK-FK references. So its result is very good.
But HGP is better than SH. HGP can analyze the co-locate relationship but SH not.
So some distributed transactions can be eliminated by these information in HGP.
Hence, HGP typically produces less number of distributed transactions than SH.


From the experimental results of the three PK-based schemes (PKH, PKR, PKRR),
we can find that the partitioning methods (hashing, range, round-robin) are not
very important for the partitioning algorithm. The three methods just got the
same results. Compared with CMRR and PKRR, we obtained that the important thing for
a partitioning scheme is the selection of partitioning key. Choosing the
right partitioning keys can get a very good result and performance. It is not so
important that which partitioning methods are chosen.

\subsubsection{TATP}

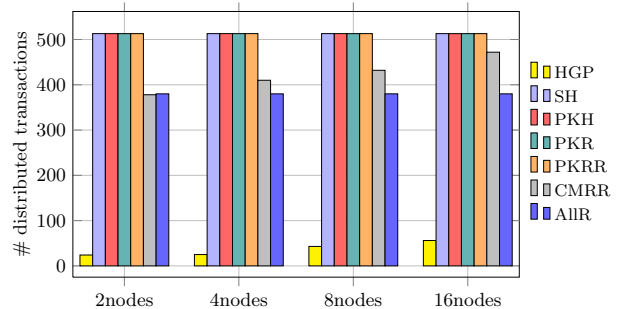
\begin{figure} \centering
\vskip -5pt
\begin{tikzpicture}[scale=0.8]
    \begin{axis}[ 
        ylabel=\# distributed transactions,
        y label style={at={(-0.08,0.50)},anchor=near ticklabel},
        height=6cm,
        width=9cm,
        grid=major,
        ybar=0pt,
        bar width=6pt,
        enlarge x limits=0.15,
        symbolic x coords={2nodes,4nodes,8nodes,16nodes},
        xtick=data,
        extra y ticks={100,300,500},
        legend style={at={(1.01,0.50)},anchor=west,legend columns=1,draw=none,font=\small},
        legend cell align=left,
        cycle list name=experimentbar,
    ]
    \addplot
    coordinates {(2nodes,24) (4nodes,25) (8nodes,43) (16nodes,56)};
    \addplot
    coordinates {(2nodes,513) (4nodes,513) (8nodes,513) (16nodes,513)};
    \addplot
    coordinates {(2nodes,513) (4nodes,513) (8nodes,513) (16nodes,513)};
    \addplot
    coordinates {(2nodes,513) (4nodes,513) (8nodes,513) (16nodes,513)};
    \addplot
    coordinates {(2nodes,513) (4nodes,513) (8nodes,513) (16nodes,513)};
    \addplot
    coordinates {(2nodes,378) (4nodes,410) (8nodes,432) (16nodes,472)};
    \addplot
    coordinates {(2nodes,380) (4nodes,380) (8nodes,380) (16nodes,380)};
    \legend{HGP,SH,PKH,PKR,PKRR,CMRR,AllR}
    \end{axis}
\end{tikzpicture}
\vskip -10pt
\caption{\# Distributed Transactions of TATP}
\label{fig:tatp}
\vskip -5pt
\end{figure}

We conducted the TATP experiments using 2000 transactions.
We partitioned the data into 2, 4, 8, 16 nodes separately.
The result is shown in Figure \ref{fig:tatp}.
%

From this experiment result, we can find that our HGP is far better than other schemes.
Its proportion of number of distributed transactions is under 5 percents,
while other schemes' proportion are greater than 20 percents.

The three PK-based schemes performed very bad in this experiments.
It performed even worse than the CMRR and AllR. It indicates
the importance of partitioning keys selection. SH is also bad since it can't choose
the suitable partitioning keys just analyzing the database schema.
The correlations between data items are not visible at the schema level in Epinions.

\subsubsection{Epinions}

\begin{figure} \centering
\vskip -5pt
\begin{tikzpicture}[scale=0.8]
    \begin{axis}[ 
        ylabel=\# distributed transactions,
        y label style={at={(-0.08,0.50)},anchor=near ticklabel},
        height=6cm,
        width=9cm,
        grid=major,
        ybar=0pt,
        bar width=6pt,
        enlarge x limits=0.15,
        symbolic x coords={2nodes,3nodes,4nodes,5nodes},
        xtick=data,
        extra y ticks={25,75,125},
        legend style={at={(1.01,0.50)},anchor=west,legend columns=1,draw=none,font=\small},
        legend cell align=left,
        cycle list name=experimentbar,
    ]
    \addplot
    coordinates {(2nodes,47) (3nodes,22) (4nodes,24) (5nodes,23)};
    \addplot
    coordinates {(2nodes,112) (3nodes,112) (4nodes,112) (5nodes,112)};
    \addplot
    coordinates {(2nodes,118) (3nodes,118) (4nodes,118) (5nodes,118)};
    \addplot
    coordinates {(2nodes,118) (3nodes,118) (4nodes,118) (5nodes,118)};
    \addplot
    coordinates {(2nodes,118) (3nodes,118) (4nodes,118) (5nodes,118)};
    \addplot
    coordinates {(2nodes,60) (3nodes,72) (4nodes,70) (5nodes,73)};
    \addplot
    coordinates {(2nodes,101) (3nodes,101) (4nodes,101) (5nodes,101)};
    \legend{HGP,SH,PKH,PKR,PKRR,CMRR,AllR}
    \end{axis}
\end{tikzpicture}
\vskip -10pt
\caption{\# Distributed Transactions of Epinions}
\label{fig:epinions}
\vskip -5pt
\end{figure}
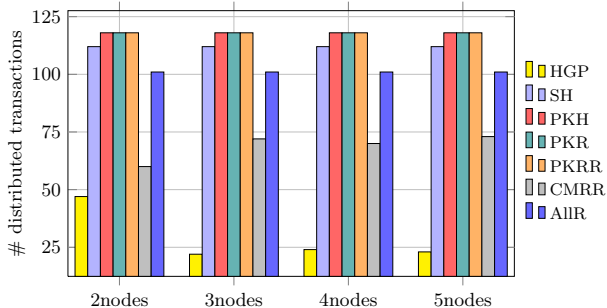

The result of the experiment using Epinions benchmark is shown in Figure \ref{fig:epinions}.
The experiment used 200 transactions. We partitioned the database into 2,3,4,5 nodes separately.
The experiment generated the similar result with TATP benchmark.

HGP is far better than other schemes. SH, PKH, PKR and PKRR generated the same number of
distributed transactions. They all choosed primary keys as the partitioning keys.
The methods (round-robin, range, hashing) used to distribute the data are not the key factor.
On the contrary, CMRR choosed the most accessed attributes as the partitioning keys. Hence, it produced
just the half number of distributed transactions of the PK ones.


\begin{figure*} \centering
\vskip -10pt
\includegraphics[height=2.2in]{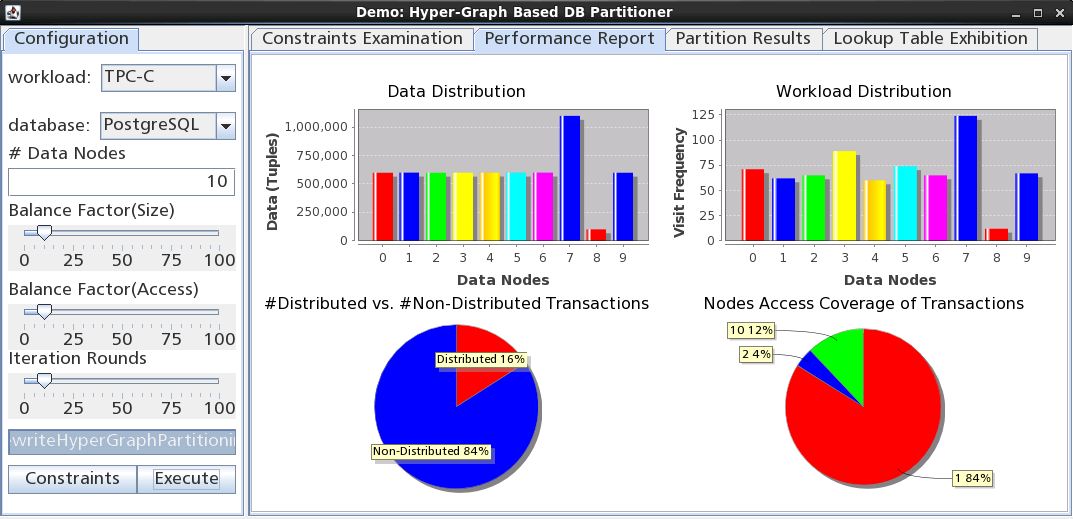}
\caption{System Demonstration}\label{fig:gui}
\vskip -10pt
\end{figure*}

\vspace{2mm}
In a conclusion, our partitioning scheme HGP performs better than other schemes if we
use the number of distributed transaction as the key metric.
Other experiments will be conducted in the near future.
Other performance metrics will be used to compare these partitioning schemes.
We also built a demo prototype, as shown in Figure \ref{fig:gui}.

\section{RELATED WORKS}
\label{sec:relatedwork}
Database partitioning is very crucial for scale transactional database,
and it is very challenging for choosing or designing the optimal partitioning
scheme for a given workload and database.

There already exist many kinds of general-purpose partitioning algorithms, among
which round-robin, range-based, hashing are the most widely used~\cite{paradb}.
These algorithms are
very effective for data analytical workload which scan very large data sets. But for
transactional workload, these methods typically produce multiple nodes access therefore
produce distributed transactions when more than one tuple is accessed in a query.

In the meantime, more ad-hoc and flexible partitioning schemes tailored for specific-
purpose applications were also developed, such as the consistent hashing of Dynamo~\cite{dynamo},
Schism~\cite{schism} and One Hop Replication~\cite{onehop}, etc..

Bubba provides many heuristic approaches
to balance the access frequency rather than the actual
number of tuples across partitions~\cite{bubba}. This algorithm
is simple and cheap, but doesn't guarantee perfect balancing of
processing.
Schism provides a novel workload aware graph-based
partitioning scheme~\cite{schism}. The scheme can get
balanced partitions and minimize the number of distributed
transactions.

Scaling social network applications has been widely reported to
be challenging due to the highly interconnected nature of the data.
One Hop Replication is an approach to scale these applications by
replicate the relationships, providing immediate access to all data
items within 'one hop' of a given record~\cite{onehop}.

~\cite{lookuptable} provides a fine-grained partitioning called lookup tables for distributed databases.
With this fine-grained partitioning, related individual tuples (e.g., cliques of friends) are co-located together in the same partition in order to reduce the number of distributed transactions. But for the tuple-level lookup table,
the database need store a large amount of meta-data about which partition each tuple resides in.
It consumes large storage space and makes the lookup operation not very efficient.

Consistent hashing~\cite{consistent_hash} can be used to minimize the data
moving when doing re-partitioning. But it may cause nonuniform
load distribution. Dynamo~\cite{dynamo} extends consistent
hashing by adding virtual nodes. It provides different
partitioning strategies on load distribution which can
ensure uniform load distribution at the same time of providing
excellent re-partitioning performance. Other works such as
CRUSH~\cite{crush} and FastScale~\cite{fastscale} can also provide algorithms
which can be used for re-partitioning.

\section{CONCLUSION}
\label{sec:conclusion}
In this technical report, we propose
a fine-grained hyper-graph based database partitioning system for
transactional workloads.
Our hyper-graph based database partitioning scheme has the following major advantages
over previous ones.

First, our scheme can reach much fine-grained and accurate partitioning results thus works well
for all kinds of transactional workloads, by taking tuple groups as the minimum
components of partitions. On the one hand, since tuple groups are directly calculated
based on the workload information, compared with blind round-robin, range
or hash partitioning methods, they are more likely to successfully cluster tuples that
will eventually be co-accessed by transactions. As a result, our approach can lead
to a fewer number of distributed transactions. On the other hand, by splitting tuple
groups into smaller ones, we can more easily mitigate the issues of data skew and
workload skew.

Second, our scheme is very light-weight and efficient. It has good scalability, as the size of the generated
hyper-graph depends only on the workload size but not on the database size. Unlike
the previous approaches, it does not need to interact with the query optimizer for
cost estimation, whose overhead is quite significant. This is feasible in practice, as
the dominant performance bottleneck of transactional workloads lies in the number
of distributed transactions, which can be directly counted from the hyper-graph
partitioning result. Moreover, the repartitioning of the hyper-graph can be done
incrementally.

Third, our scheme is very flexible. The users are allowed to input their performance
expectations at the beginning. During the partitioning iterations, They can watch
the visualized partitioning effects, and optionally provide feedback based on his expertise
and domain knowledge so as to affect the decision on whether the system
should proceed to do graph refinement and re-partitioning, as well as to provide
suggestions on how the current hyper-graph should be refined. With such interactions
with the users, our approach is able to reach configurable and precise balance
between the partitioning speed and partitioning quality.

\eat{
We detailed describe the partition model, system architecture and
some experiments. Through modeling the database partitioning problem
as a k-way multi-constraints hyper-graph partitioning problem and deriving a
min-cut of the hyper-graph, our system can minimize the total number
of distributed transactions in the workload, balance the sizes
and workload accesses of the partitions and satisfy all the partition
constraints imposed. The partitioning scheme proposed is very light-weight, efficient and flexible.
It can reach much fine-grained and accurate partitioning results that can works well
for all kinds of transactional workloads. The experiments show that
our partitioning scheme produces far less number of distributed transactions than other schemes.
}

\bibliographystyle{splncs}

\end{document}